\RequirePackage{rotating}
\documentclass[fleqn,usenatbib]{mnras}

\usepackage{newtxtext,newtxmath}

\usepackage[T1]{fontenc}

\DeclareRobustCommand{\VAN}[3]{#2}
\let\VANthebibliography\thebibliography
\def\thebibliography{\DeclareRobustCommand{\VAN}[3]{##3}\VANthebibliography}

\usepackage{graphicx}	
\usepackage{amsmath}	
\usepackage{graphicx}
\usepackage{epstopdf}
\usepackage{xcolor}
\usepackage{times}
\usepackage{ulem}
\usepackage{rotating}
\usepackage{multirow,bigdelim}
\usepackage{adjustbox}
\usepackage{pdflscape}
\usepackage{soul}

\definecolor{sbc}{rgb}{0.1, 0.6, 0.1}
\definecolor{check}{rgb}{0.7, 0.1, 0.5}

\newcommand{\gtsim}{$\buildrel > \over \sim $}

\def\f404{$4_{04}-3_{03}$}
\def\f505{$5_{05}-4_{04}$} 
\def\cm3{cm$^{-3}$} 

\title[HCN in comet 3I/ATLAS]{JCMT detection of HCN emission from 3I/ATLAS at 2.1 AU}

 \author[I. M. Coulson et al.]{  
Iain M. Coulson,$^{1}$\thanks{E-mail: i.coulson@eaobservatory.org}
Yi-Jehng Kuan,$^{2,3}$\thanks{E-mail: kuan@ntnu.edu.tw}
 Steven B. Charnley,$^{4}$
Martin A. Cordiner,$^{4,5}$
Yo-Ling Chuang,$^{3,2}$
 \newauthor Yueh-Ning Lee,$^{2}$
Min-Kai Lin,$^{3}$
Stefanie N. Milam,$^{4}$
Bannawit Pimpanuwat,$^{6}$
Nathan X. Roth,$^{4,7}$
Micha{\l} {\.Z}{\'o}{\l}towski$^{8}$
\\
$^{1}$East Asian Observatory, 660 N.A'ohoku Place, Hilo, Hawaii 96720, USA\\
$^{2}$Center of Astronomy and Gravitation and Department of Earth Sciences, National Taiwan Normal University,  88, Sec. 4, Ting-Chou Road, Taipei 11677, Taiwan, ROC.\\
 $^{3}$Academia Sinica Institute of Astronomy \& Astrophysics,	P. O. Box 23-141, Taipei 106, Taiwan, ROC\\
$^{4}$Astrochemistry Laboratory, Code 691, NASA Goddard Space Flight Center, 8800 Greenbelt Rd., Greenbelt, Maryland 20771, USA\\
$^{5}$Department of Physics, Catholic University of America, Washington, DC 20064, USA\\
$^{6}$National Astronomical Research Institute of Thailand (NARIT),
260 Moo 4 Donkaew, Maerim, Chiang Mai 50180, Thailand\\
$^{7}$Department of Physics, American University, 4400 Massachusetts Ave NW, Washington, DC, USA \\
$^{8}$Faculty of Chemistry, Nicolaus Copernicus University in Torun, 7 Gagarin Street, 87-100 Torun, Poland\\
}

\date{Accepted XXX. Received YYY; in original form ZZZ}

\pubyear{2025}

\begin{document}
\label{firstpage}
\pagerange{\pageref{firstpage}--\pageref{lastpage}}
\maketitle

\begin{abstract}
 We report the detection of HCN($J=3-2$) rotational emission from comet 3I/ATLAS at a
heliocentric distance of 2.13 AU with the James Clerk Maxwell Telescope (JCMT). 
Observations were conducted between  07 August 2025 and 14 September 2025 (UT) using the $^{'}$$\overline{U}$$^{'}$$\overline{u}$  heterodyne receiver and ACSIS spectroscopic backend. 
  The HCN line was  detected at $> 6 \sigma$  on 14 Sep 2025 (UT) and a production rate of $Q({\rm HCN})=(4.0\pm1.7)\times10^{25}$~s$^{-1}$ was derived by non-LTE radiative  transfer modelling.  
Preliminary estimates of the HCN/H$_2$O and CN/HCN abundance ratios suggest values similar to Solar System comets.
\end{abstract}



 
\section{Introduction}   \label{sec:intro}  
 

 The discovery of the first two interstellar objects (ISOs) ---  1I/'Oumuamua and 2I/Borisov ---  has ushered forth an exciting new development in astronomy:  the detailed study of the Galactic population of small bodies  (Jewitt \& Seligman 2022). 
Interstellar comets, such as 2I/Borisov, most probably formed in a similar manner as those in the Solar System and, as such,  they can provide crucial  information on the early evolution of the protostellar disk in which they formed and also, perhaps,  on their potential for initiating planetary prebiotic chemistry in these systems, as has been proposed for the early Earth (Chyba \& Hand 2005).  Studies of ISOs may also provide detailed insights into the physical and chemical processing of icy small bodies (i.e. comets)  during the billions of years they spend traversing the interstellar medium (Seligman et al. 2022). 
 

We now have the rare opportunity to observe the third recorded interstellar visitor to our Solar System:  comet 3I/ATLAS\footnote{Also designated comet C/2025 N1 (ATLAS)} (Seligman et al. 2025).
 The discovery of 3I  (at 4.5 AU)  provides a unique opportunity for detailed  study of an ISO,  allowing comparisons to be made with the apparitions of Solar System comets.  With a proposed origin  from the Galactic thick
disk (Hopkins et al. 2025),  and an age of $\sim 3-11$ Gyr (Taylor \& Seligman 2025), comet 3I may exhibit a distinctive   chemical composition indicative of  sub-solar metallicity and an extended sojourn in the Galactic interstellar medium significantly longer than the age of the Solar System. 



To date, molecular observations of 3I have been reported when the comet was at heliocentric distances of  $R_{\rm h}$ \gtsim 2.9AU. 
 Of the coma daughter molecules typically observed at optical  wavelengths:   OH, CN, C$_2$, C$_3$, NH and NH$_2$ (Biver et al. 2024)  only 2 have been detected  :  OH (at UV wavelengths, Xing et al. 2025) and CN,   with derived upper limits 
 pointing to particularly strong depletion of  C$_2$ and C$_3$ (Schleicher 2025; Rahatgaonkar et al. 2025; Salazar Manzano et al 2025).
Interstellar comet 2I/Borisov was found to be rich in CO compared to most Solar System comets (Cordiner et al 2020; Bodewits et al. 2020) and was probably also rich in CO$_2$ (Opitom et al 2021; McKay et al 2024).  
Infrared observations of 3I/ATLAS at $R_{\rm h} \sim 3.3$AU, with the James Webb Space Telescope (JWST) and with SPHEREx,  detected the major parent molecules emanating directly from the nucleus - CO$_2$, CO, H$_2$O - and indicate that  CO$_2$ molecules dominate the coma composition (Cordiner  et al. 2025; Lisse et al. 2025). 
However, as yet,  there have been no reports of other less abundant parent compounds that can be detected in both the IR and mm/submm, e.g. HCN, CH$_3$OH, H$_2$CO, NH$_3$, CS.   
  Of these, due to its large dipole moment, HCN exhibits strong emission in the mm/submm spectral region making it the most easily detected parent molecule  and therefore a marker for the onset of significant sublimation of nuclear ices. 
  Chemically, HCN measurements can place constraints on the presence of additional sources for CN in the coma, other than from HCN photolysis (via the CN/HCN ratio), and, in principle, on the C and N isotopic ratios (Biver et al. 2024).
  
We have been observing 3I/ATLAS through August-September 2025 in a program to track the development of outgassing activity through  mm/submm emissions, particularly from CO and  HCN.   In this Letter we report the detection of HCN in the coma of 3I/ATLAS when the comet  was inbound and 2.1 AU from the Sun. 

\section{Observations}  \label{sec:obs}

We observed comet 3I in August and September 2025 using the   James Clerk Maxwell Telescope (JCMT). 
 The JCMT is located on  Maunakea, Hawaii, at an
altitude of 4092 m. It has a 15m diameter primary mirror and operates at frequencies (wavelengths)
between 220 and 690 GHz (1.3 and 0.4 mm).   
 We obtained spectra of what 
is usually the strongest mm-wave cometary line accessible 
in the  A-band, namely the $J=3-2$ transition of HCN 
at 265.8864~GHz.  The HCN$(J=3-2)$ and HCN$(J=4-3)$ transitions are typically employed to monitor cometary outgassing activity and their emission can be routinely detected in bright comets by JCMT (e.g. Coulson et al. 2017, 2020). 
Observations were carried out using the single-receptor A-band receiver 
$^{'}$$\overline{U}$$^{'}$$\overline{u}$  
with its output passed to the ACSIS digital autocorrelation spectroscopic
backend. ACSIS is configurable to provide a multiplicity of
spectral resolutions and bandpass widths. Maximum spectral resolution 
of 31 kHz ($\sim$0.03~km~s$^{-1}$) over a bandpass of 250~MHz is desirable 
for observations of single, narrow lines such as the one observed here.
 
\begin{table*} 
\caption{Observations of the HCN($J=3-2$) line emission from comet 3I/ATLAS}
\begin{center}
\begin{tabular}{ccccccc}
\hline
\hline
UT date  & Days before & R$_{h}$ & $\Delta$ & Integration& Integrated    & Q(HCN)        \\
 & perihelion &         &          & Time &  Intensity     &              \\
         &            & (AU)     & (AU)      & (hrs.) & (mK~km~s$^{-1}$) & (10$^{25}$~mol.~s$^{-1}$) \\
\hline
 20250807.25  &  -83.24  & 3.290  & 2.734  & 5.0  & $<$5.5   & $<$1.5  \\
  20250815.25  & -75.24 & 3.002  & 2.653  & 3.1 &  $<$9.6  & $<$2.1  \\
   20250822.25  & -68.24  & 2.814 &  2.615  & 2.5 &  $<$9.6  & $<$1.7  \\
   20250828.25   &-62.24  & 2.629  & 2.588  & 2.5 &  $<$9.0  & $<$1.4   \\
   20250903.25   &-56.24  & 2.448  & 2.569 &  2.0 &  $<$9.2 &  $<$1.1 \\
20250907.25  & -52.24 & 2.331  & 2.559 & 1.5 & 12.1 \underline{+} 3.9  & 1.2 \underline{+}  0.5 \\
   20250914.25  & -45.24  & 2.133 & 2.544 & 1.6 &   36.9  \underline{+}  6.0  &   {4.0  \underline{+}   1.7}    \\
 \hline
  \end{tabular}  
\end{center}
  \label{tab:HCN32obs} 
     \end{table*}

Pointing of the telescope is checked routinely by observations of 
astronomical point sources. The all-sky pointing precision of JCMT is
$\sim$2$\arcsec$ rms in each of the (azimuth, elevation) coordinates,
although, once pointed, telescope tracking accuracy over the course of 
an hour is better than 1$\arcsec$. The quality of pointing during our 
observing run was perfectly nominal. The FWHM beam at 265.8864 GHz is 18.4". 
Pointing and tracking of the nucleus of comet 3I/ATLAS was accomplished using 
ephemerides $\#$25 \& $\#$26 from JPL/Horizons. Implementation of these JPL 
ephemerides at JCMT was found to replicate the predicted coordinates to 
better than 1$\arcsec$ on all nights. 

The observations here were made in {\it stare} mode, in which the 
telescope tracks the coordinates of the target during the ON-SOURCE phase
and moves to a nearby {\it sky} position in the OFF-SOURCE phase, with spectra
being accumulated during each phase. The relevant subtraction of spectra is 
performed by ACSIS software.
We employed {\it beam-switching} (BMSW) between the ON and OFF phases, which is 
implemented by oscillating the hyperbolic telescope secondary mirror at 1~Hz. 
We used a BMSW throw of 90$\arcsec$ which
provided a sky position about 164,000~km from the comet during this period;
sufficiently far so as to avoid contamination of the spectrum of the sky by 
emission from the coma.

JCMT heterodyne observing does not feature continuous Doppler tracking,
so individual observation lengths were kept to 30 minutes or less, 
keeping Doppler smearing to less than 0.02~km~s$^{-1}$.
JCMT observational data files are automatically complemented by records of 
the opacity of the atmosphere (tau) as measured by the JCMT Water Vapor Meter (WVM). 
The WVM intercepts part of the JCMT beam, and records the opacity at a rate 
of 1~Hz. These measures are used in the routine data processing in order to 
correct instrumental signal levels for atmospheric extinction. Our data
were all collected when tau, the zenith opacity at 225~GHz, was less than 0.15~(nepers), and as 
good as 0.06 on the two days of observing in September. 

Routine reduction of the spectroscopic data used {\it Starlink} software
and followed the methodology of Thomas \& Currie (2016). The data were also 
routinely adjusted onto the cometocentric velocity frame in order to facilitate 
aggregation of data; the velocity information for this being provided by the 
JPL/Horizons ephemeris.

Calibration of the measured intensities was performed by spectroscopic 
observations of astronomical standards, usually one measure every 2 or 3 hours,
or before and after the observing sequence if shorter than 2~hours.
The measured integrated intensities of those standards were found 
to deviate by at most a few percent from their catalogued values,
so our measures of 3I/ATLAS are effectively on the JCMT T$_{a}$$^{*}$ 
intensity scale.

\begin{figure}
\centering 
\includegraphics[width=3.5in, height=1.8in]{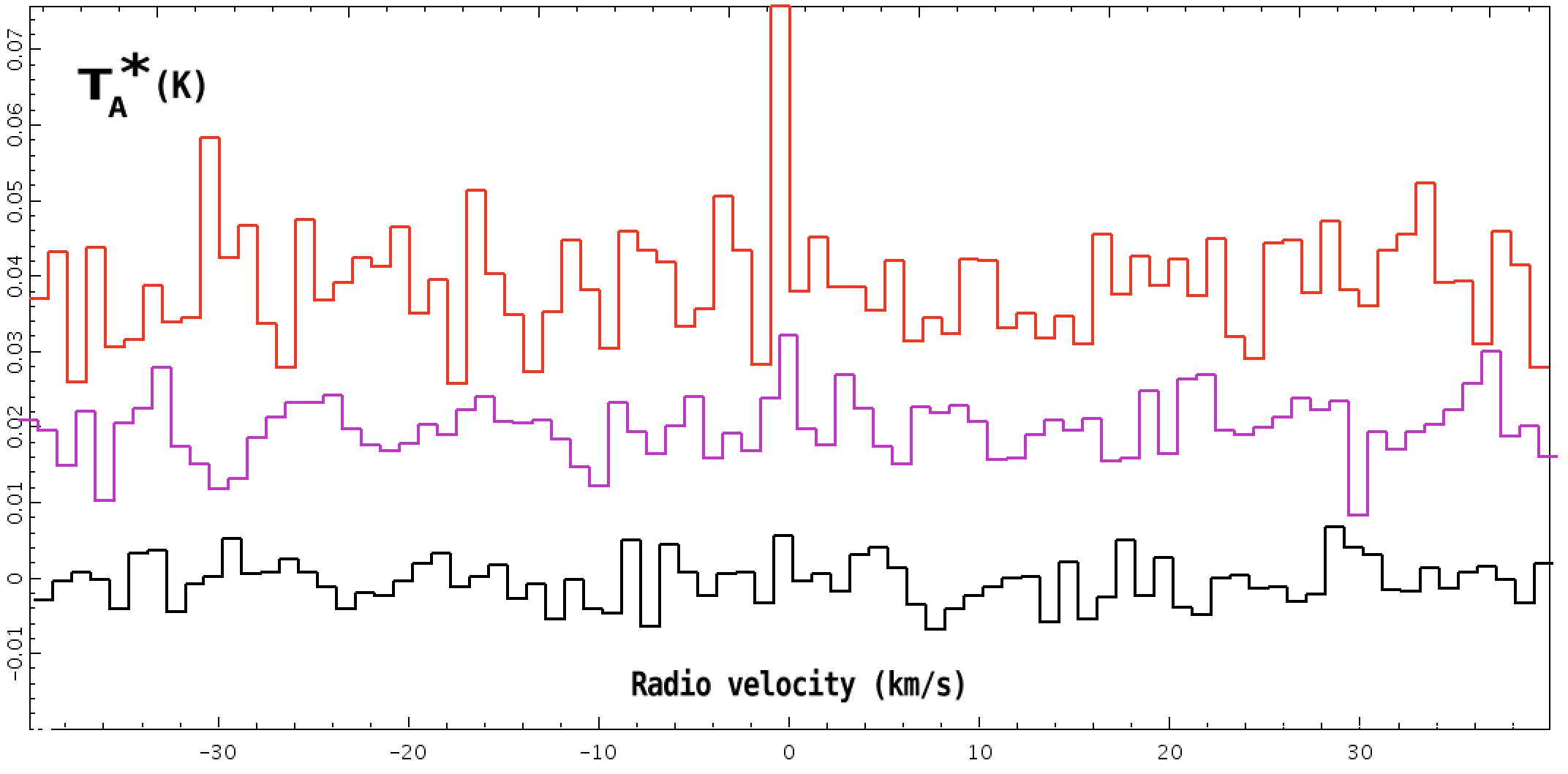}
\caption{HCN($J=3-2$) spectra of comet 3I/ATLAS obtained at JCMT:
 UT 03 Sep 2025 (bottom, black), UT 07 Sep 2025 (middle, purple), UT 14 Sep 2025 (top, red). 
Data are shown
in the cometocentric velocity frame at 1.0~km~s$^{-1}$ resolution.   
 } \label{fig:HCN32spec}
\end{figure}

Results from our HCN($J=3-2$) observations of  comet 3I  are given in 
{\color{blue} Table \ref{tab:HCN32obs}} and   {\color{blue} Figure \ref{fig:HCN32spec}}  shows the spectra of the detected lines.
Columns 1-4 give the UT date of the mid-point of each observation 
(yyyymmdd.dd), 
the interval in days prior to  perihelion, the  heliocentric distance 
and the geocentric distance, R$_{h}$ and $\Delta$ (AU), 
respectively.  Column 5  gives the combined integration time (on + off, hours) 
and column 6  gives the  integrated intensity of the HCN($J=3-2$) line  
(mK km/s) or the 3-sigma upper limit. The intensities measured on Sep~07 
and 14 use the peak of the spectral line at cometocentric velocity $\sim$0~km~s$^{-1}$
when displayed at a channel width of 1~km~s$^{-1}$, and the error tabulated is the 1-sigma
noise in the spectral baseline.
The statistical significance of the Sep 7 and Sep 14 detections are $ 3.1 \sigma$ and  $ 6.1 \sigma$, respectively.

%

\section{HCN profile modelling}  \label{sec:HCNprof}
 
\begin{figure}

\centering

\includegraphics[width=3.0in, height=2.4in]{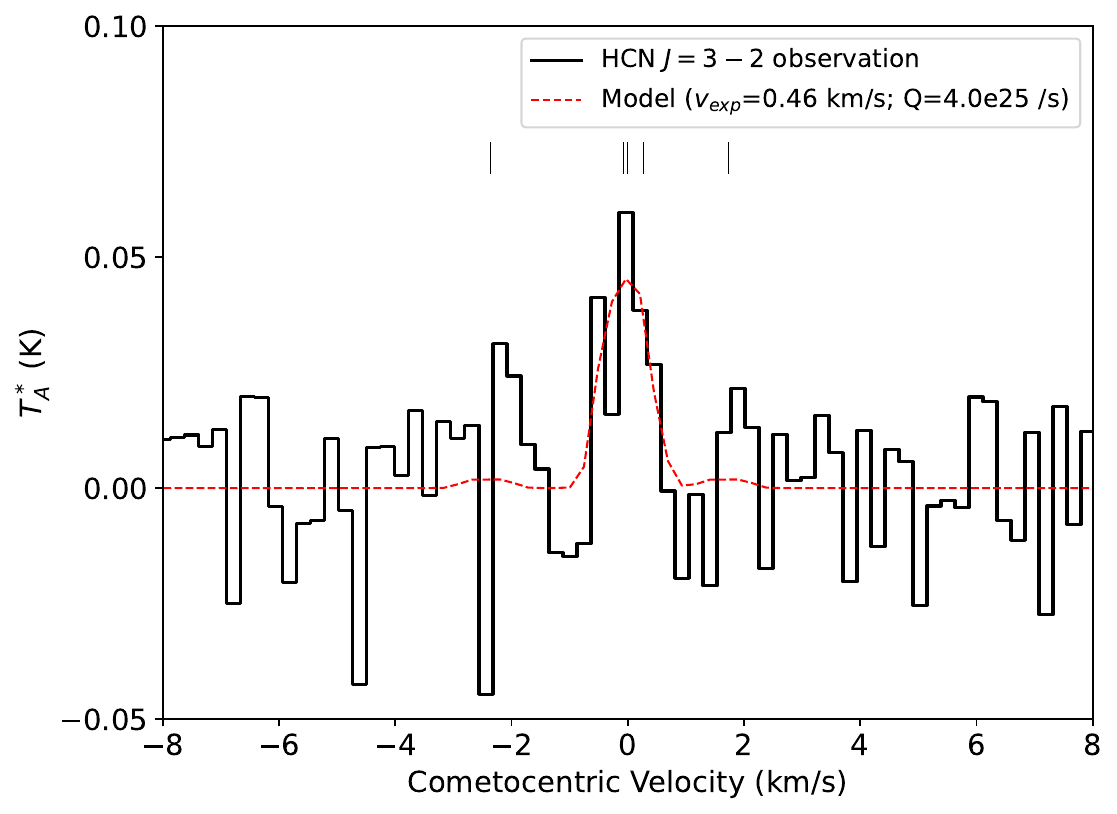}

\caption{
SUBLIME model fit to the HCN($J=3-2$)  line observed on September 14.  Here, the spectrum is shown at a 0.24 km/s channel binning, with the strongest 5 HCN hyperfine components labeled by tick marks.
Details of the line profile fitting are given in the text.
}\label{fig:SUBLIMEDFIT}

\end{figure}

  In {\color{blue} Table \ref{tab:HCN32obs}}  column 7 gives the HCN  production rate, $Q$(HCN), or the 3-sigma upper limit, in the case of no clear detection.

The observed JCMT spectra were modeled using the 1D version of the SUBLIME non-LTE radiative transfer code (Cordiner et al. 2022), to derive the HCN production rates of 3I/ATLAS, assuming spherically-symmetric sublimation, direct from the nucleus. The model includes HCN photolysis at the Huebner \& Mukherjee (2015) active Sun rate, and rotational excitation due to collisions with H$_2$O, based on the latest state-specific rate coefficients calculated by {\.Z}{\'o}{\l}towski\ et al. (2025), as well as collisions with electrons (following the method of Biver et al. 1999, with an electron density scaling factor of 0.2), and pumping by solar radiation (Crovisier \& Encrenaz 1983).

 For the Sep 14 data, for which HCN was detected at $6.1\sigma$ confidence, we performed a full spectral modeling (at 0.24 $\rm km~s^{-1}$ binning) to determine the HCN production rate and outflow velocity. A water production rate of $2\times10^{28}~s^{-1}$ was adopted for this date based upon measurements of OH obtained using the Swift satellite on 2025 September 13-14 (D. Bodwewits, private communication, 2025) and with an assumed H$_2$O ortho-to-para ratio of 3. The coma kinetic temperature was assumed to be 35$\pm15$ K, based on prior observations of comets at similar heliocentric distances ($R_{\rm h}\sim2$~AU);  see Cordiner et al. (2020) and references therein. A JCMT main beam efficiency factor of 0.66 was used and the resulting fit is shown in {\color{blue} Figure \ref{fig:SUBLIMEDFIT}}. The model spectrum, which included the hyperfine components of the HCN $J=3-2$ transition, was optimized using the LMFIT routine (Newville et al. 2014), to obtain the best fit to the observations (see {\color{blue} Figure \ref{fig:SUBLIMEDFIT}}). The retrieved HCN production rate is $Q({\rm HCN})=(4.0\pm1.7)\times10^{25}$~s$^{-1}$. The retrieved HCN outflow velocity (derived from the spectrally resolved HCN line shape) is $0.46\pm0.14 \rm ~km ~ s^{-1}$, which accounts for the HCN hyperfine components as well as the finite channel binning of the observations. The overall Doppler shift of the HCN line was allowed to vary during fitting, resulting in a best fitting radial velocity offset of $-0.06 \pm 0.10$~km\,s$^{-1}$.

 For the other dates, the HCN production rates and upper limits were derived by adopting the same outflow velocity, but with $Q(\rm H_2O)$ determined using from a logarithmic fit to the $Q(\rm H_2O)$ values as a function of $R_{\rm h}$, from Cordiner et al. (2025), Xing et al. (2025) and Bodewits et al. (private communication). In the case of non-detections, the ($3\sigma$) HCN production rate upper limits were obtained by scaling the (optically thin) model $Q({\rm HCN})$ values for each date by the ratio of the noise to model intensities, spectrally-integrated over the full-width of the model line.


\section{Conclusions} 

We report the first detection of the ubiquitous cometary molecule HCN in the interstellar comet 3I/ATLAS. 
The  derived production rate at a heliocentric distance of 2.1 AU is $Q({\rm HCN})=(4.0\pm1.7)\times10^{25}$~s$^{-1}$ and the corresponding abundance relative to H$_2$O is  $(2.0\pm0.8)\times10^{-3}$, which is comparable to the highest ratios  found in Solar System comets (HCN/H$_2$O $\approx 0.03 - 0.4\%$;  Biver et al. 2024). 

There are no available contemporaneous measurements of CO that would enable a comparison of $Q({\rm CO})/Q({\rm HCN})$ between 3I, 2I/Borisov, and other comets. For the $Q({\rm CN})/Q({\rm HCN})$ ratio we can, however, make an estimate based on published $Q({\rm CN})$ measurements when $R_{\rm h}$ was 2.94 AU and 2.85 AU (Salazar Manzano et al.  2025;  Rahatgaonkar et al. 2025)\footnote{ CN was first detected in mid-August, about one month prior to our HCN detection. The non-detection of HCN during the month between and also earlier in mid-July (see Hinkle et al. 2025) was mainly an effect of beam dilution, as at a distance of $\approx 2.7$ AU from Earth in mid-August, the linear scale of the JCMT beam at 265.9 GHz was $\approx 36,000$ km; and at 2.5 AU on Sep 14, it was $\approx 33,300$ km, compared with $\approx 2,000$ and 1,800 km for the arcsec-resolution optical telescope used for CN detection. }. 
 Employing the scaling proposed by Rahatgaonkar et al. (2025) of $Q({\rm CN}) \propto R_{\rm h}^{-9.4}$ yields $Q({\rm CN}) \approx (0.72- 2.2) \times10^{25}$~s$^{-1}$, and is consistent with a direct measurement  contemporaneous with Sep 14 of $Q({\rm CN}) \approx 1.8 \times10^{25}$~s$^{-1}$ (Hutsemekers et al. 2025). 
Thus, we find  $Q({\rm CN})/Q({\rm HCN})$$\approx 0.1-1$, indicating that HCN photolysis alone could account for the observed CN abundance (cf. Dello Russo et al. 2016).

The apparition of 3I/ATLAS  presents an extremely rare  opportunity in cometary science. Ongoing monitoring and analysis  will further reveal  the chemistry and  physical structure of this comet from another solar system.

\section*{Acknowledgements }

 
 We thank the staff of the JCMT for their support during the observing run in August and September 2025. 
We are grateful to Dennis Bodewits for sharing of observational results from the Neil Gehrels-Swift Observatory prior to publication.

These observations were obtained by the James Clerk Maxwell Telescope, operated by the East Asian Observatory on behalf of Academia Sinica Institute of Astronomy and Astrophysics and the National Astronomical Research Institute of Thailand. Additional funding support is provided by the Science and Technology Facilities Council of the United Kingdom and participating universities and organizations in the United Kingdom and Canada.

Y.-J.K. acknowledges support by the Ministry of Education, Taiwan, ROC. 
S.B.C, M.A.C., N.X.R. and S.N.M. acknowledge support by the NASA Planetary Science Division Internal 
Scientist Funding Program through the Fundamental Laboratory Research work package (FLaRe).   
M.A.C. was also supported by the National Science Foundation (grant AST-1614471). 
B.P. is supported by the Fundamental Fund of Thailand Science Research and Innovation (TSRI) through the National Astronomical Research Institute of Thailand (Public Organization) (FFB680072/0269).
M.Z. is grateful for the support of the National Science Centre (NCN)
through Grant "Sonatina 8" No. 2024/52/C/ST9/00110.

\section*{Data Availability}
The observations were made for Program S25BP001, and can be obtained after the proprietary period via the JCMT archive at
\url{http://www.cadc-ccda.hia-iha.nrc-cnrc.gc.ca/en/jcmt/}.




\begin{thebibliography}{}

 
\bibitem[Biver et al.(1999)]{1999AJ....118.1850B} Biver, N., Bockel{\'e}e-Morvan, D., Crovisier, J., et al.\ 1999, \aj, 118, 4, 1850.  
\bibitem[Biver et al.(2024)]{2024come.book..459B} Biver, N., Dello Russo, N., Opitom, C., et al.\ 2024, Comets III. 
Edited by Karen J. Meech et al., Space Science Series, University of Arizona Press, 459 
\bibitem[Bodewits et al.(2020)]{2020NatAs...4..867B} Bodewits, D., Noonan, J.~W., Feldman, P.~D., et al.\ 2020, Nature Astronomy, 4, 867.  
\bibitem[Chyba \& Hand (2005)]{2005ARA&A..43...31C} Chyba~C.F. \& Hand~K.P. 2005, \araa,  43, 31
\bibitem[Cordiner et al.(2020)]{2020NatAs...4..861C} Cordiner, M.~A., Milam, S.~N., Biver, N., et al.\ 2020, Nature Astronomy, 4, 861.  
\bibitem[Cordiner et al.(2022)]{2022ApJ...929...38C} Cordiner, M.~A., Coulson, I.~M., Garcia-Berrios, E., et al.\ 2022, \apj, 929, 1, 38 

\bibitem[Cordiner et al.(2025)]{2025ApJ...991L..43C} Cordiner, M.~A., Roth, N.~X., Kelley, M.~S.~P., et al.\ 2025, \apjl, 991, 2, L43
 \bibitem[Coulson et al.(2017)]{2017AJ....153..169C} Coulson I.M. et al.\ 2017 \aj, 153, 169
\bibitem[Coulson et al.(2020)]{2020AJ....160..182C} Coulson, I.~M., Liu, F.-C., Cordiner, M.~A., et al.\ 2020, \aj, 160, 4, 182

\bibitem[Crovisier \& Encrenaz(1983)]{1983A&A...126..170C} Crovisier, J. \& Encrenaz, T.\ 1983, \aap, 126, 1, 170. 
\bibitem[Dello Russo et al.(2016)]{2016Icar..278..301D} Dello Russo, N., Kawakita, H., Vervack, R.~J., et al.\ 2016, \icarus, 278, 301.  
\bibitem[Feinstein et al.(2025)]{2025ApJ...991L...2F} Feinstein, A.~D., Noonan, J.~W., \& Seligman, D.~Z.\ 2025, \apjl, 991, 1, L2.  
 \bibitem[Hinkle et al.(2025)]{2025arXiv251202106H} Hinkle, J.~T., Yang, B., Meech, K.~J., et al.\ 2025, , arXiv:2512.02106 
\bibitem[Hopkins et al.(2025)]{2025ApJ...990L..30H} Hopkins, M.~J., Dorsey, R.~C., Forbes, J.~C., et al.\ 2025, \apjl, 990, 2, L30.  
\bibitem[Huebner \& Mukherjee(2015)]{2015P&SS..106...11H]} Huebner W.F. \& Mukherjee J.\ 2015, P\&SS 106, 11
\bibitem[Hutsem{\'e}kers et al.(2025)]{2025arXiv250926053H} Hutsem{\'e}kers, D., Manfroid, J., Jehin, E., et al.\ 2025, , arXiv:2509.26053. 
\bibitem[Jewitt \& Seligman(2023)]{2023ARA&A..61..197J} Jewitt, D. \& Seligman, D.~Z.\ 2023, \araa, 61, 197.  
\bibitem[Lisse et al.(2025)]{2025RNAAS...9..242L} Lisse, C.~M., Bach, Y.~P., Bryan, S., et al.\ 2025, RNAAS, 9, 9, 242.  
\bibitem[McKay et al.(2024)]{2024DPS....5630101M} McKay, A., Opitom, C., Jehin, E., et al.\ 2024, DPS, 56, 301.01. 
\bibitem[Mumma \& Charnley(2011)]{2011ARAA..49..471M} Mumma M.~J. \& Charnley S.~B.\ 2011, \araa, 49, 471
\bibitem[Newville et al.(2014)]{2014zndo.....11813N} Newville, M., Stensitzki, T., Allen, D.~B., et al.\ 2014, Zenodo, 0.8.0.  
\bibitem[Opitom et al.(2021)]{2021A&A...650L..19O} Opitom, C., Jehin, E., Hutsem{\'e}kers, D., et al.\ 2021, \aap, 650, L19.  
\bibitem[Rahatgaonkar et al.(2025)]{2025arXiv250818382R} Rahatgaonkar, R., Carvajal, J.~P., Puzia, T.~H., et al.\ 2025, , arXiv:2508.18382.  
\bibitem[Salazar Manzano et al.(2025)]{2025ApJ...993L..23S} Salazar Manzano, L.~E., Lin, H.~W., Taylor, A.~G., et al.\ 2025, \apjl, 993, 1, L23 
\bibitem[Seligman et al.(2022)]{2022PSJ.....3..150S} Seligman, D.~Z., Rogers, L.~A., Cabot, S.~H.~C., et al.\ 2022, PSJ,  3, 7, 150.  
\bibitem[Seligman et al.(2025)]{2025ApJ...989L..36S} Seligman, D.~Z., Micheli, M., Farnocchia, D., et al.\ 2025, \apjl, 989, 2, L36.  
\bibitem[Schleicher(2025)]{2025ATel17352....1S} Schleicher, D.\ 2025, The Astronomer's Telegram, 17352, 1. 
\bibitem[Taylor \& Seligman(2025)]{2025ApJ...990L..14T} Taylor, A.~G. \& Seligman, D.~Z.\ 2025, \apjl, 990, 1, L14.  
\bibitem[Thomas \& Currie(2016)]{}Thomas H.~S. \& Currie M.~J., 2016, \\ https://starlink.eao.hawaii.edu/devdocs/sc20.htx/sc20.html

\bibitem[Xing et al.(2025)]{2025ApJ...991L..50X} Xing, Z., Oset, S., Noonan, J., et al.\ 2025, \apjl, 991, 2, L50  

\bibitem[{\.Z}{\'o}{\l}towski et al.(2025)]{2025MNRAS.540..626Z} {\.Z}{\'o}{\l}towski, M., Lique, F., {\.Z}uchowski, P., et al.\ 2025, \mnras, 540, 1, 626.  

\end{thebibliography}



\bsp	
\label{lastpage}
\end{document}